\documentclass[5p,times]{elsarticle}

\usepackage{amssymb}
\usepackage{amsmath}

\usepackage{cancel}

\usepackage{orcidlink}

\usepackage{graphicx} 
\usepackage[utf8]{inputenc}
\usepackage[T1]{fontenc}

\usepackage{xspace}

\usepackage{natbib}

\usepackage{color}
\usepackage{xcolor}
\usepackage{comment}
\usepackage{amsmath}
\usepackage{booktabs}

\usepackage{listings}
\usepackage{adjustbox}

\usepackage{acronym}

\usepackage{blindtext}
\usepackage{url}
\usepackage{hyperref}

\usepackage[normalem]{ulem}

\usepackage[switch]{lineno} 
\runningpagewiselinenumbers

\newcommand{\code}[1]{\lstinline!#1!\xspace}

\journal{Nucl. Instr. \& Methods in Physics Research, A}

\begin{document}

\begin{frontmatter}

\title{Efficient and precise Cherenkov-based charged particle timing using SiPMs}

\author[a]{M.~N.~Mazziotta\,\corref{mycorrespondingauthor}\,\orcidlink{0000-0001-9325-4672}\,}
\cortext[mycorrespondingauthor]{Corresponding author}
\ead{mazziotta@ba.infn.it}

\author[b]{A.~Di~Mauro\,\orcidlink{0000-0003-0348-092X}\,}

\author[a,c]{M.~Giliberti\,\orcidlink{0009-0007-2835-2963}\,}

\author[a,c]{A.~Liguori\,\orcidlink{0009-0001-4240-6362}\,}

\author[a]{L.~Lorusso\,\orcidlink{0000-0002-2549-4401}\,}

\author[a]{E.~Nappi\,\orcidlink{0000-0003-2080-9010}\,}

\author[a,b,c]{N.~Nicassio\,\orcidlink{0000-0002-7839-2951}\,}

\author[a]{G.~Panzarini\,\orcidlink{0000-0002-2586-1021}\,}

\author[a]{R.~Pillera\,\orcidlink{0000-0003-3808-963X}\,}

\author[a,c]{G.~Volpe\,\orcidlink{0000-0002-2921-2475}}

\affiliation[a]{organization={Istituto Nazionale di Fisica Nucleare, Sezione di Bari}, addressline={via Orabona 4}, city={Bari}, postcode={I-70126}, country={Italy}}

\affiliation[b]{organization={CERN, the European Organization for Nuclear Research}, addressline={Esplanade des Particules 1}, postcode={1211}, city={Geneva}, country={Switzerland}}

\affiliation[c]{organization={Dipartimento di Fisica ``M. Merlin'' dell'Universit\`a  e del Politecnico di Bari}, addressline={via Amendola 173}, city={Bari}, postcode={I-70126}, country={Italy}}

\begin{abstract}
Dedicated R\&D efforts are currently underway to couple a thin Cherenkov radiator to Silicon Photomultiplier (SiPM) arrays for precise charged particle Time-of-Flight (ToF) measurements. The prompt nature of Cherenkov radiation makes it an ideal candidate for achieving ultimate timing performance in a ToF detector. Using a thin radiator with a high refractive index, such as fused silica, enables the generation of a fast signal from charged particles that exceed the Cherenkov threshold.
A crucial requirement for approaching the target time resolution is the optimization of both the radiator material and thickness, as well as the optical coupling to the SiPM arrays. In this work, we present the main factors that affect the time resolution and the expected performance achieved through a detailed Monte Carlo simulation and the comparison with beam test results.
\end{abstract}

\end{frontmatter}

\section{Introduction}
\label{sec:intro}

The detection of the Cherenkov light is not only a   powerful technique for identifying charged particles over a large momentum range, as exploited by Ring-Imaging Cherenkov (RICH) detectors, but it results to be a very effective way to enhance the time resolution in Time-of-Flight (TOF) devices because of its promptness nature.
When charged particles exceed the Cherenkov threshold and traverse a thin radiator slab made of a high refractive index material, such as synthetic quartz (fused silica), the resulting photons can generate a fast, well-timed signal in photon sensors like Microchannel Plate Photomultipliers (MCP-PMTs)(see, e.g., Ref.s~\cite{Credo:2004qgy,Inami:2006cp,Krizan:2008zz,Vavra:2009qpk,Albrow:2012ha}).

A very promising  step forward in the development of Cherenkov-based TOFs is the replacement of MCP-PMTs with Silicon Photomultipliers (SiPMs). In fact, SiPMs are progressively being recognized as ideal sensors for precision timing measurements, owing to their excellent performance in key metrics. These metrics include time resolution, high photon detection efficiency (PDE), insensitivity to magnetic fields, and a low material budget (see, e.g.,~\cite{Gundacker:2020cnv}). The Single Photon Time Resolution (SPTR) of a SiPM is typically $\sigma_{\text{SPTR}} \sim 100\,\text{ps (RMS)}$ or better, and it generally improves with increasing excess bias voltage. 
Furthermore, the SPTR improves when photons are detected in the central region of the Single-Photon Avalanche Diodes (SPADs) that constitute the SiPM microcells~\cite{Acerbi:2019qgp}.

In the following sections, we present the main factors that affect the time resolution and the expected performance achieved through a Monte Carlo simulation are presented and discussed.

\section{SiPM-based charged particle timing measurements}
\label{sec:SiPMtime}

\begin{figure}[!th]
 \centering
 \includegraphics[scale=0.35]{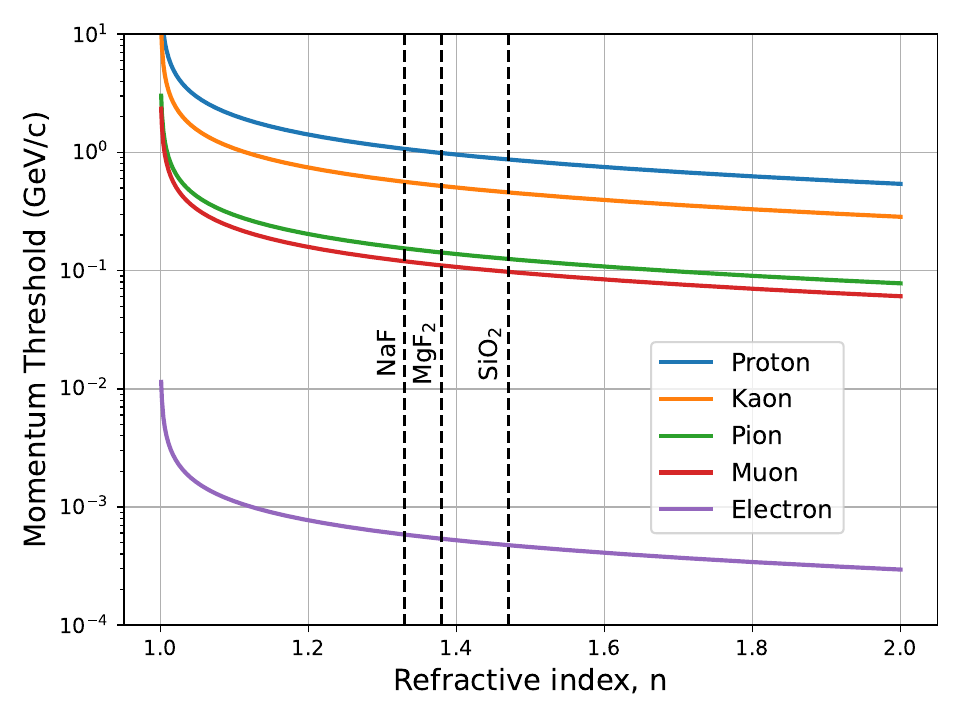}
 \caption{Cherenkov momentum threshold as a function of the refractive index $n$ for various charged particle species (electrons, muons, pions, kaons, protons). The vertical dashed lines indicate the refractive indices of common radiator materials: NaF, $\text{MgF}_2$, and fused silica ($\text{SiO}_2$) at $\approx$ 400 nm.}
 \label{fig:pth}
\end{figure}

Charged particles impinging on a high refractive index material (radiator) can produce Cherenkov radiation even at low momenta. Figure~\ref{fig:pth} shows the momentum thresholds for different particle species as a function of the refractive index $n$. For radiators with $n > 1.3$, the Cherenkov momentum threshold is below $1\,\text{GeV}/c$ for protons and even lower for lighter particles. Cherenkov photons are emitted in the forward direction at an angle $\theta_C$ with respect to the particle direction, given by the relation $\cos \theta_C = 1/n \beta$, where $\beta$ is the particle speed normalized to the speed of light in vacuum and $n=n(\lambda)$.

Efficient detection of Cherenkov photons requires a radiator with high transparency for wavelengths exceeding the near ultraviolet (NUV) band. In addition, to minimize reflection losses caused by refractive index mismatch at the boundaries, the photodetector must be optically well coupled to the radiator. This is typically achieved using an optical cement or glue with a refractive index close to that of the radiator.

\begin{figure}[!h]
    \centering
    \includegraphics[scale=0.3]{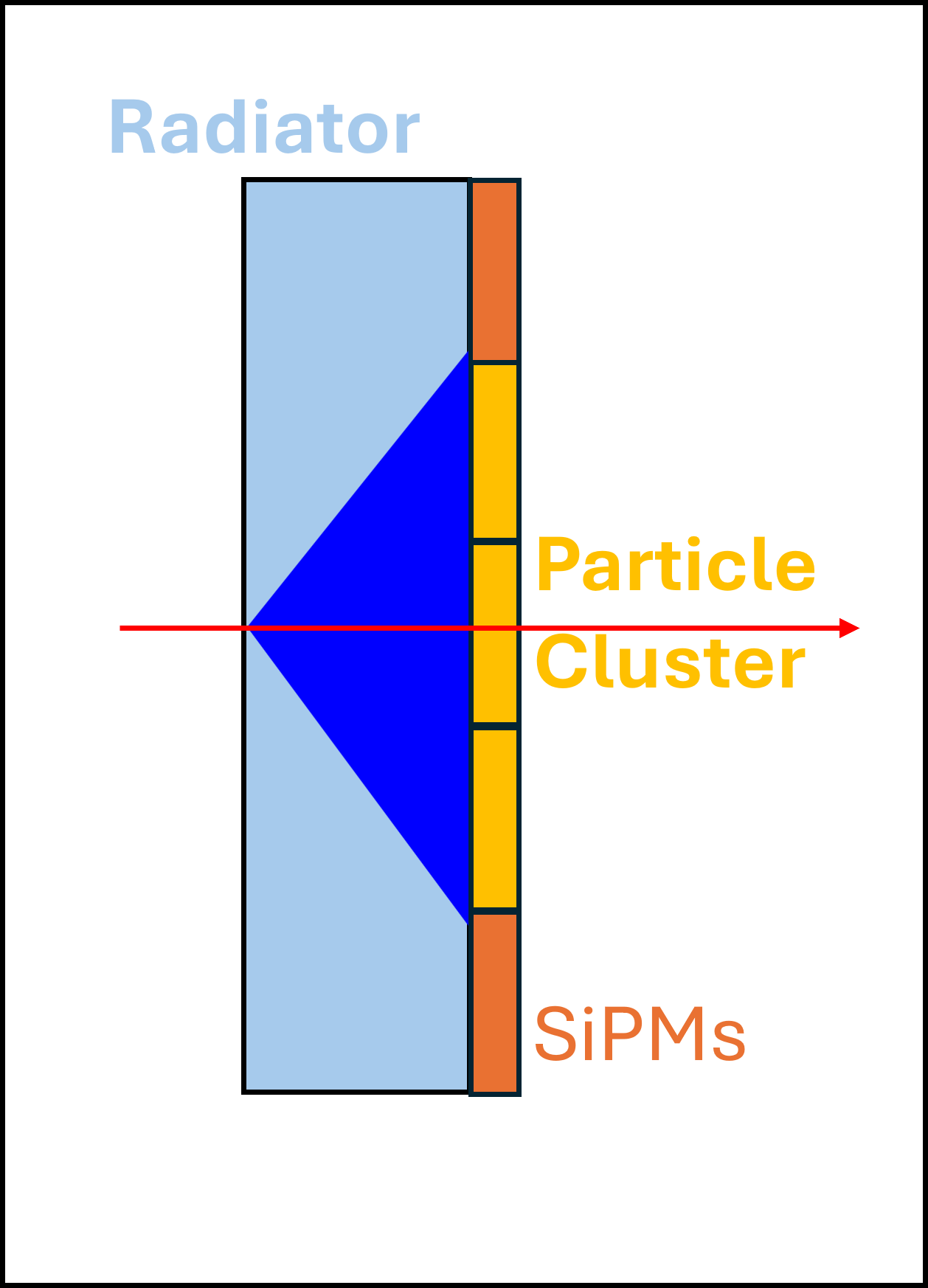}
    \caption{Conceptual sketch depicting the optical coupling of a thin Cherenkov radiator with a SiPM array. The blue shaded area represents the possible trajectories of Cherenkov photons generated along the path of the incident charged particle. The yellow  pixels indicate the SiPMs fired by the Cherenkov photons, neglecting possible  reflections.}
    \label{fig:ChRadSiPM}
\end{figure}

A Cherenkov radiator, typically a few mm thick with an optimized refractive index, can be optically coupled to a SiPM array to yield  precise timing signals~\cite{Carnesecchi:2023dfq,Nicassio:2023kux,Mazziotta:2025kdr,Mazziotta:2025zxj,particles8040094}. Figure~\ref{fig:ChRadSiPM} shows a conceptual sketch of a SiPM array coupled with a thin radiator. A charged particle generates a narrow cluster of high-amplitude hits ("track hits") localized near its path. The track position can be precisely reconstructed, even with a fine SiPM pixel pitch, by accounting for the charge information from each fired pixel. Furthermore, the overall charged particle detection efficiency is expected to be close to $100\%$, effectively mitigating the impact of the dead areas between adjacent SiPM cells.

The time resolution benefits of the high number of photoelectrons ($N_{pe}$) expected in each readout channel. 
Specifically, if $N_{\text{pixel}}$ is the number of fired pixels used for the time measurement, the system time resolution $(\sigma_{\text{system}})$ is expected to improve proportionally to $1/\sqrt{N_{\text{pixel}}}$ (assuming Gaussian distribution), relative to the single SiPM channel time resolution. However, charge sharing across multiple pixels can negatively impact on the time resolution, particularly when only a few PEs are collected by peripheral pixels, leading to an increased contribution from the electronic time jitter.

Indeed, the sketch shown in Fig.~\ref{fig:ChRadSiPM} involves multiple optical interfaces, including the radiator-SiPM optical glue, the SiPM coating resin, the anti-reflection coating (ARC), the SiPM passivation layer, and the silicon substrate. The optical couplings must therefore be carefully optimized to suppress reflections and prevent photon losses and to mitigate signal delays.

\section{Main factors affecting the time resolution}

The time resolution of a SiPM-radiator system is affected by several factors. These include: the spread in photon arrival times due to the Cherenkov photon travel paths from their emission points to the SiPM surface; the chromatic dispersion, where the refractive index dependency on photon wavelength broadens the distribution of the Cherenkov emission angle and the resulting arrival time distribution; multiple reflections off radiator interfaces before detection, which might increase the photon transit time. Detector-specific limitations are the SiPM intrinsic time resolution, which encompasses photoelectron creation, multiplication jitter, and charge transit time spread, and the electronic time jitter, caused by noise-related fluctuations in the readout circuitry.

The primary function of SiPMs is photon detection. However, an ionization signal is simultaneously generated when a charged particle traverses the SiPM depleted region. This ionization creates electron-hole pairs, resulting in a prompt electrical signal alongside the optical signal (see, e.g.,~\cite{Brigida:2004ff,mazziotta2008,Riegler:2021xtq}). 
This secondary signal component contributes to the overall timing and charge reconstruction.

In a detector geometry similar to the one shown in Fig.~\ref{fig:ChRadSiPM}, and assuming normal incidence of a charged particle while neglecting secondary effects (e.g. scattering, and photon reflections), the arrival time of a Cherenkov photon produced at a position $x$ along the radiator (of thickness $d$) is given by:

\begin{equation}
t(x) = \frac{x}{\beta\,c} + \frac{\left( d-x \right)\,n(\lambda)}{c\, \cos{\theta_C}}.
\label{eq:ph_toa}
\end{equation}
The first term in the r.h.s. of Eq.~\ref{eq:ph_toa} is the ToF of the charged particle to the emission point $x$, while the second term is the ToF of the Cherenkov photon from $x$ to the photodetector surface. The maximum time spread between photons produced at the radiator entrance ($x=0$) and those at the exit ($x=d$) is:

\begin{equation}
 \Delta t_{\text{max}}=\frac{d}{\beta\,c} \left( \beta^2 n^2(\lambda) - 1 \right) \propto d.
 \label{eq:dtmax}
\end{equation}
This spread is linearly proportional to the radiator thickness. Assuming a uniform distribution of photon production points, the corresponding time resolution ($\sigma_t$) due to this spread is:

\begin{equation}
  \sigma_t(d) = \frac{\Delta t_{\text{max}}}{\sqrt{12}} \propto \frac{d}{\sqrt{12}}.
\end{equation}

The intrinsic time resolution from the SiPM ($\sigma_{pe}$) depends statistically on the number of photoelectrons ($N_{pe}$). Assuming independent arrival times that are averaged, $\sigma_{pe}$ scales inversely with the square root of $N_{pe}$ and can be expressed as:

\begin{equation}
 \sigma_{pe} \approx \frac{\sigma_{SPTR}}{\sqrt{N_{pe}}} \propto \frac{1}{\sqrt{d}}.
\end{equation}
Since $N_{\text{pe}}$ is generally proportional to the radiator thickness ($N_{\text{pe}} \propto d$), the intrinsic detector time resolution can also be related to the radiator thickness as $\sigma_{\text{pe}} \propto 1/\sqrt{d}$.

Finally, the overall electronic contribution ($\sigma_{\text{ele}}$) primarily originates from the front-end (FE) amplification and the subsequent time-to-digital conversion (TDC). These terms are combined in quadrature: $\sigma^2_{\text{ele}} = \sigma^2_{\text{FE}} + \sigma^2_{\text{TDC}}$. The FE time jitter is typically noise-dominated and scales inversely with $N_{pe}$ (i.e. with the charge): $\sigma_{\text{FE}} \propto 1/N_{\text{pe}}$, showing that noise is significantly reduced with a higher photon yield. The TDC jitter is dictated by the intrinsic resolution, related to the converter's Least Significant Bit (LSB) and is given by $\sigma_{\text{TDC}}=\text{LSB}/\sqrt{12}$. 

\begin{figure}[!th]
 \centering
 \includegraphics[scale=0.35]{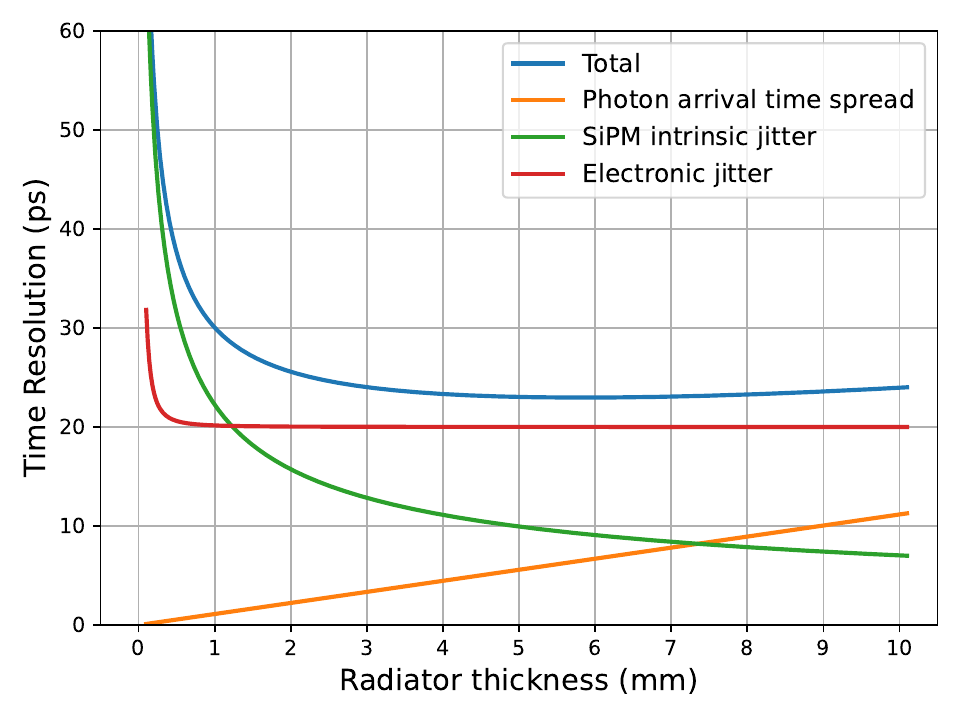}
 \caption{Expected time resolution as a function of a fused silica radiator thickness. Colored lines show the different contributions. The maximum spread of photon arrival time is calculated according Eq.\ref{eq:dtmax}  weighted with the Cherenkov photons emission spectrum in the wavelength range from 260 nm to 900 nm and the SiPM PDE.}
 \label{fig:sigt_d}
\end{figure}

Figure~\ref{fig:sigt_d} presents the resulting time resolution as a function of the radiator thickness, calculated using reference parameters: $\beta=1$; Cherenkov photons emission in the wavelength interval from 260 nm to 900 nm (Eq. 34.44 of Ref. \cite{ParticleDataGroup:2024cfk}); fused silica radiator including the wavelength dependence of the refractive index ~\cite{1965JOSA...55.1205M,TAN1998158}; SiPMs with SPAD pitch of $75 \mu m$, a peak PDE of 50\% at the nominal operational bias  voltage and T=25$^\circ$C \cite{s13660-3075CS} and $\sigma_{\text{SPTR}} = 100\,\text{ps}$; electronic jitter modeled as $\sigma_{\text{ele}} = 50\,\text{ps}/N_{\text{pe}} \oplus 20\,\text{ps}$. The results show that the expected total time resolution is below $30\,\text{ps}$ for radiator thicknesses $d \ge 1 \,\text{mm}$ with the considered SIPM. This dependence illustrates the design trade-off: a thinner radiator reduces the geometric time spread ($\sigma_t \propto d$), but concurrently reduces the number of collected photons, which degrades $\sigma_{\text{pe}} \propto 1/\sqrt{d}$ and increases the $N_{\text{pe}}$-dependent electronic jitter. We remark here that the limit of the time resolution is due to the performance of the read-out electronic implemented in the simulation.

\section{Time resolution Monte Carlo calculation}
\label{sec:MC}

We employ a Monte Carlo method to calculate the expected time resolution for a SiPM array with pixelated read-out. First, a charged particle track, assumed to be on-axis, is simulated as it traverses a detector region that covers multiple pixels. Cherenkov photons are subsequently generated uniformly along this track.  
The average number of Cherenkov photons produced per unit track length are calculated from 260 nm to 900 nm corresponding to the typical SiPM PDE spectrum.
For each event, the number of Cherenkov photons is calculated with a Poisson distribution according to the average number and then we track the path of each photon with a given wavelength $\lambda$ generated according to the Cherenkov spectrum. The collected photons are converted to $N_{pe}$ based on the specific SiPM PDE. 

The time-of-arrival of these photons at the SiPM surface is calculated using Eq.\ref{eq:ph_toa}. For each read-out channel, the $N_{pe}$ is determined by considering the incident photons and the SiPM's PDE. The final step involves calculating the overall time resolution by incorporating the influence of the intrinsic SiPM jitter and the electronic time resolution, both of which are parameterized by the calculated $N_{pe}$ values and modeled using a Gaussian probability distribution.
We simulated three configurations: i) 3 mm SiPMs ($75 \, \mu\text{m}$ microcell); ii) 2 mm SiPMs ($50 \, \mu\text{m}$ microcell); iii) 1 mm SiPMs ($50 \, \mu\text{m}$ microcell). 
The PDE as a function of the wavelenght for the considered SiPMs has been taken from Ref.\cite{s13660-3075CS} at the typical operational voltage and T=25$^\circ$C. 

\begin{figure}[!th]
    \centering
    \includegraphics[scale=0.45]{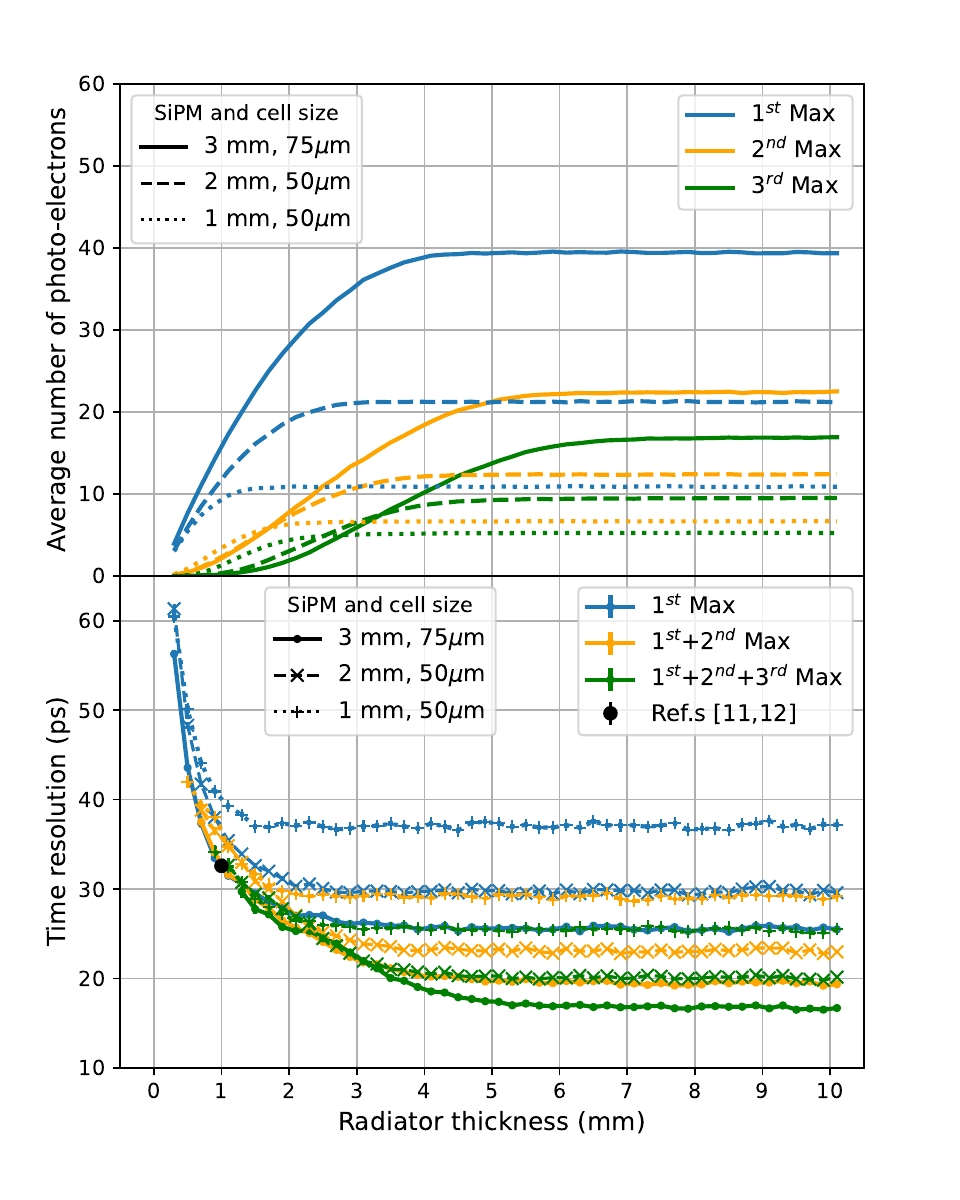}
    \caption{Top panel: collected photoelectrons in the three highest-hit pixels as a function of the fused silica radiator thickness. Bottom panel: time resolution as a function of the fused silica radiator thickness, calculated using the three channels with the highest observed charge. Three SiPM configurations are considered: (i) 3 mm ($75 \, \mu\text{m}$ microcell); (ii) 2 mm ($50 \, \mu\text{m}$ microcell); and (iii) 1 mm ($50 \, \mu\text{m}$ microcell). The black filled circle represent the single SiPM resolution derived from the experimental results reported in Refs.~\cite{Mazziotta:2025zxj,particles8040094} (see text for details).}
    \label{fig:npe_sgtmax_d}
\end{figure}

The top panel of Fig.~\ref{fig:npe_sgtmax_d} shows the values of $N_{pe}$ in the three highest-hit pixels  as a function of the fused silica radiator thickness. The number of photoelectrons in the highest-charge pixel increases with the radiator thickness and subsequently saturates at different values: $\approx 4$ mm for the 3 mm SiPM configuration, $\approx 2.5$ mm for the 2 mm SiPM configuration, and $\approx 1.3$ mm for the 1 mm SiPM configuration. A similar trend is observed in the pixels recording the second and third highest charge, though saturation is shifted toward higher radiator thicknesses. This saturation behavior is driven by a combination of the Cherenkov angle and the longitudinal production of photons along the charged track. As expected, $N_{pe}$ in these channels is higher for the 3 mm SiPM compared to the 2 mm and 1 mm configurations. Furthermore, the maximum-$N_{pe}$ channel consistently records a number of photoelectrons that is approximately a factor 2 higher than the second and third highest-$N_{pe}$ channels. 

The bottom panel of Fig.~\ref{fig:npe_sgtmax_d} shows the expected time resolution for the three simulated configurations as a function of the fused silica radiator thickness. This resolution was calculated in three ways: first, using only the maximum-$N_{pe}$ channel, and then by averaging the arrival time across the two and three channels with the highest observed number of photoelectrons. As anticipated, the time resolution is primarily driven by the channels with the highest observed charge. The configuration with 3 mm SiPMs yields a superior resolution even with thin radiators. The time resolution improves when averaging across two or three channels with the highest observed $N_{pe}$, as more photons are collected in these channels; however, this improvement needs the use of a thicker radiator.

In Refs.~\cite{Mazziotta:2025zxj,particles8040094}, a time resolution of approximately $46 \text{ ps}$ is reported. This value was evaluated by comparing the highest-charge channels from two SiPM arrays with 2 mm and 3 mm pixel sizes, both coupled to a 1 mm thick fused silica radiator. Assuming equal contributions from these sensors, the corresponding time resolution for a single SiPM is $\approx32.5 \text{ ps}$ ($46 \text{ ps}/\sqrt{2}$). 
This single-sensor value is also shown in the bottom panel of Fig.~\ref{fig:npe_sgtmax_d} with a black filled circle. With the parameters discussed in this paper, the calculated time resolution for a 1 mm radiator thickness is approximately $30 \text{ ps}$, which is consistent with the data reported in Refs.~\cite{Mazziotta:2025zxj,particles8040094}. For a more rigorous comparison, however, the exact electronic parameters and the SiPM time resolution from the reference experiments should be incorporated into the simulation.

\section{Conclusions}
\label{sec:con}

In this study, we calculated the expected time resolution of a SiPM-based ToF detector for charged particles traversing high refractive index materials that produce Cherenkov photons.
Our Monte Carlo simulation, based on fundamental assumptions, allowed us to effectively discuss the main parameters affecting the time resolution and evaluate the expected performance of different detector configurations. While our calculated results show good consistency with current experimental data, further studies and optimizations are needed to approach the ultimate timing performance of such a detector.

Specifically, future optimizations should account for multiple reflections that occur due to the presence of various surfaces and material interfaces (e.g., the radiator-SiPM resin interface, anti-reflection coatings, passivation layers, and the silicon surface).
In particular, during photon propagation, other characteristic optical angles, such as the total internal reflection angle and Brewster’s angle, should also be precisely incorporated into the simulation. Finally, to accurately reflect real-world performance, the SiPM electronic signal should be simulated including the associated electronics jitter.

The intrinsic capabilities of a SiPM-based ToF detector make it highly suitable for integration with RICH detectors by sharing the same photodetector layer, leading to a compact and efficient configuration~\cite{Nicassio:2023kux, Mazziotta:2025kdr, Mazziotta:2025zxj, Pillera:2025iqx}.

\bibliographystyle{elsarticle-num-names}
\bibliography{Mazziotta_RICH2025.bib}

\end{document}